\documentclass[twocolumn,prd,floatfix,showpacs,showkeys]{revtex4}
\usepackage[latin9]{inputenc}
\usepackage{float}
\usepackage{units}
\usepackage{textcomp}
\usepackage{amsmath}
\usepackage{amsthm}
\usepackage{graphicx}
\usepackage{esint}

\usepackage{slashed}
\usepackage{physics}
\usepackage{braket}

\makeatother

\begin{document}
\title{Critical chiral hypersurface of the magnetized NJL model}
\author{Angelo Mart\'inez and Alfredo Raya}
\affiliation{Instituto de F\'{\i}sica y Matem\'aticas, Universidad Michoacana de San Nicol\'as de Hidalgo. Edificio C-3, Ciudad Universitaria. Francisco J. M\'ujica s/n, Col. Fel\'icitas del R\'{\i}o, C.P. 58040, Morelia, Michoac\'an, Mexico.}

\begin{abstract}
    In pursuit of sketching the effective magnetized QCD phase diagram, we find conditions on the  critical coupling for chiral symmetry breaking in the Nambu--Jona-Lasinio model in a nontrivial thermo-magnetic environment. Critical values for the plasma parameters, namely, temperature and magnetic field strength for this to happen are hence found in the mean field limit.
    The magnetized phase diagram is drawn from the criticality condition for different models of the effective coupling describing the inverse magnetic catalysis effect. 
\end{abstract}

\maketitle

\section{Introduction}

Understanding the behavior of quantum chromodynamics (QCD)  in a  nontrivial thermo-magnetic environment of quarks and gluons is both a very hot topic and a hard nut to crack (a recent review on the sketch of the magnetized QCD phase diagram can be found in Ref.~\cite{review}). These extreme  conditions are met, for instance, in compact stars  and in peripheral relativistic heavy ion collisions, which give rise to  magnetic fields of around $m_\pi^2$ at RHIC and $15m_\pi^2$ at LHC~\cite{skokov09}. Although these fields are short-lived~\cite{skokov2}, in order to study its effect on the chiral transition, a common starting point is to regard them  as uniform in space and time in such a manner that the scenario for the said transition is impacted by the magnetic field strength in a non-trivial manner as the heat bath temperature is increased. In this view, in an attempt to sketch the magnetized QCD phase diagram, after neglecting density effects, lattice QCD simulations available in Refs.~\cite{bali:2012a,bali:2012b,bali:2014} reveal, on the one hand, that the chiral condensate grows with the magnetic field strength for temperatures below $T_0$, the pseudocritical temperature for the chiral transition in absence of the magnetic field,  in accordance with the universal phenomenon of magnetic catalysis (MC)~(see Ref.~\cite{shovkovy} for a  review). On the other hand, 

 for $T\simeq T_0$, a turnover behavior settles and the pseudocritical transition temperature decreases as the strength of the magnetic field increases. This phenomenon has been dubbed  Inverse Magnetic Catalysis (IMC) effect. A  plausible explanation for this phenomenon is that the strong QCD coupling exhibits a non-trivial thermomagnetic behavior such that the competition between the magnetic field strength and temperature renders the coupling to reach its asymptotically free limit in an accelerated manner.

Establishing the detailed properties of the QCD coupling as a function of the magnetic field strength is a formidable task, mostly because the effects of the field belong to the non-perturbative domain, where the strong coupling is less known~(see, for instance,  Refs.~\cite{review,miranskyreview,shov}). Deriving the properties of the coupling constant from vertex corrections in a weak magnetic field has been recently done in Ref.~\cite{ayala-vertex}, for instance. Nevertheless, effective models  might still be able to capture general features of the running of the coupling with $B$. 

Within the Linear Sigma models (LSM),  the full thermomagnetic dependence of 
the self-coupling~\cite{ayalaetal:LSM1} as well as the
quadratic and quartic couplings~\cite{ayalaetal:LSM2} have been obtained including the medium screening effects properly. These findings capture the basic traits of the IMC effect, namely, decreasing of the couplings with increasing magnetic field for $T>T_0$ as well as the critical temperatures for chiral symmetry restoration. Without incorporating those effects, it is not possible to reproduce the growth and decreasing of the chiral condensate with the magnetic field strength for temperatures smaller and larger than $T_0$, respectively.

Nambu--Jona-Lasinio (NJL) model~\cite{NJL}~(see Refs.~\cite{klevansky,buballa} for reviews), its extensions~\cite{PNJL} and non-local variants~\cite{nNJL} have widely been used to capture some non-perturbative features of QCD. The local model is non-renormalizable and the coupling needs to exceed a critical value in order to be able to describe chiral symmetry breaking. Furthermore, in the mean field limit, a medium-independent coupling constant fails to incorporate important dynamics to be able to reproduce the traits of IMC. A non-trivial dependence of the plasma parameters for the coupling is required to explain the said phenomenon. Based on lattice simulations, there have been attempts to describe a nontrivial thermomagnetic behavior of the coupling in the local NJL model~\cite{pade,aftab,allofus1,allofus2,gastao,costa}. 
For the non-local extensions, early works~\cite{marcelonnjl} establish that for weak magnetic fields, IMC effects are not observed when the coupling is considered in its mean field limit. For strong magnetic fields, however, this type of models have recently been observed to  offer a natural explanation to the IMC effect~\cite{norberto} in the sense that the way form factors depend on the magnetic field correspond to a backreaction effect of the sea quarks on the gluon fields that makes the interaction strength decrease as the magnetic field strength increases. Furthermore, these models provide clues for the simultaneity of the chiral and confinement/deconfinement transitions when models are coupled to a Polyakov loop potential. 

In  the present article we  study the boundary of the critical chiral hypersurface in parameter space of the magetized NJL model  to determine the conditions on the coupling constant for which chiral symmetry breaking takes place  in terms of  the plasma parameters, namely, temperature and magnetic field strength.
In the limiting case of a medium independent coupling constant, we are able to identify the critical values of the temperature and magnetic field strength enough to break chiral symmetry. Some indirect hints of IMC are explicitly observed even in this regime insofar as an entirely different analytic behavior of the coupling as a function of the magnetic field strength for temperatures above and below $T_0$ develops. We further explore the magnetized phase diagram along this critical chiral hypersurface assuming a dressing  of the coupling by the plasma
as is modeled to describe IMC in Refs.~\cite{pade,aftab,allofus1,allofus2,gastao,costa}. For this purpose, we have organized the remaining of the article as follows: In Sec.~\ref{sec:gap} we derive the details of the gap equation for the magnetized NJL model within the proper-time regularization scheme. Section~\ref{sec:surf} is devoted to obtain the critical hypersurface for chiral symmetry breaking of the model in parameter space. A criticality condition among the parameters of the medium is established that divides symmetric from asymmetric domains in parameter space according to chiral symmetry. The phase diagram obtained from this criticality condition  is sketched in Sec.~\ref{sec:comp} for several proposals of the coupling which are dressed by the medium in accordance with the IMC effect. Concluding remarks are presented in Sect.~\ref{sec:concl}.

\section{Gap Equation}\label{sec:gap}

\begin{figure}
\begin{centering}
\includegraphics[width=\linewidth]{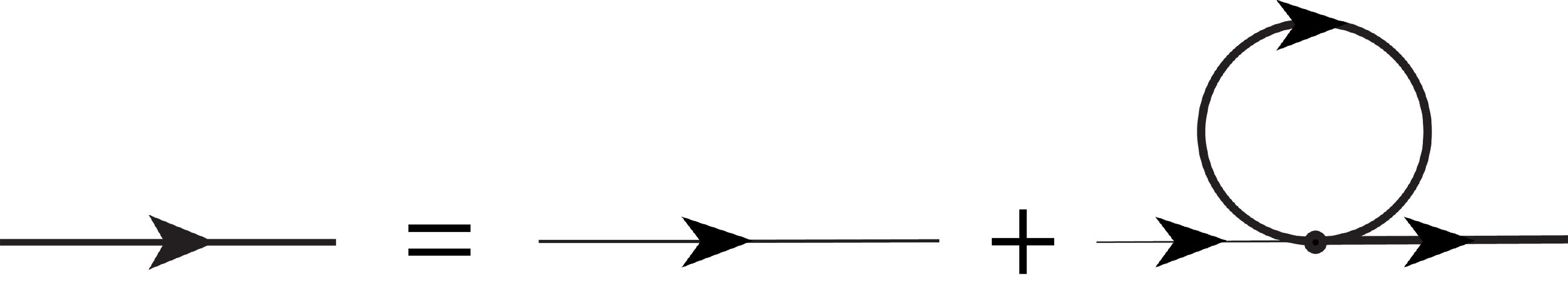}
\par\end{centering}
\caption{\label{fig:Schwinger-Dyson-equation}Schwinger-Dyson equation corresponding
to the Hartree-Fock approximation.}
\end{figure}

One of the first successful attempts to describe strong interaction is the model put forward by Nambu and Jona-Lasinio in analogy with superconductivity~\cite{NJL}. Such a model is described through the Lagrangian
\begin{equation}
    {\cal L}=\bar\psi \left( i {\not \! \partial} -m_q\right) \psi + G\left[ (\bar\psi\psi)^2+(\bar\psi i\gamma_5\vec\tau\psi)^2\right]\;,
\end{equation}
where the fields $\psi$ are in modern literature regarded as quark fields with current mass $m_q$ and $G$ is the (dimensionful) coupling of the model. Here, $\vec\tau$ correspond to the Pauli matrices acting on isospin space. 
Within the Hartree-Fock approximation,
the corresponding Schwinger-Dyson (gap) equation for the quark propagator is pictorially described in Fig.~\ref{fig:Schwinger-Dyson-equation}, 
which can be expressed as
\begin{equation}
m=m_{q}-2G\braket{\bar{\psi}\psi},\label{eq:gapgral}
\end{equation}
where $m$ is the dynamically generated mass and $-\braket{\overline{\psi}\psi}$ is the chiral
condensate, defined as, 
\begin{equation}
-\braket{\bar{\psi}\psi}=\int\frac{d^4p}{(2\pi)^4}{\rm Tr}\left[iS\left(p\right)\right].\label{eq:condensate}
\end{equation}
Here,
\begin{equation}
 S(p)=\frac{1}{{\not \! p}-m}
\end{equation}
is the dressed quark propagator with $m$ the dynamically generated mass, which is momentum independent. Even though we consider isospin symmetric light quark flavors, in what follows we consider the gap equation for a single light quark flavor and set $N_f=1$. The trace, however, runs over Dirac and color spaces.  

As mentioned before, the model is non-renormalizable, and therefore, integrals must be regulated.
We adopt the proper-time regularization scheme~\cite{schwinger}, which allows to incorporate magnetic field effects straightforward. We follow the procedure first used in Refs.~\cite{allofus1,allofus2}, which we briefly describe in here.

Considering a uniform magnetic field of strength $B$ aligned with the third spatial axis, the translationally invariant part of the quark propagator adopts its  
 Schwinger representation~\cite{schwinger}
\begin{align}
S(p) & =-i\int_{\Lambda}^{\infty}\frac{ds}{\cos(q_{f}Bs)}e^{is\left(p_{\parallel}^{2}-p_{\perp}^{2}\frac{\tan(q_{f}Bs)}{q_{f}Bs}-m^{2}\right)}\nonumber \\
 & \bigg\{(\cos(q_{f}Bs)+\gamma_{1}\gamma_{2}\sin(q_{f}Bs))(m+\slashed p_{\parallel})\nonumber \\
 & -\frac{\slashed p_{\perp}}{\cos(q_{f}Bs)}\bigg\},\label{eq:prop}
\end{align}
were 

$q_{f}$ is the absolute value of the quark charge and $p_{\parallel}$ and $p_{\perp}$ are the parallel and perpendicular components of the four-momentum, defined as:
\begin{eqnarray}
p_{\parallel}^{\mu} & =&\left(p_{0},0,0,p_{3}\right)\;,\nonumber \\
p_{\perp}^{\mu} & =&\left(0,p_{1},p_{2},0\right),
\end{eqnarray}
and $\Lambda$ is the proper time cut-off, which has canonical dimensions of $[{\rm Mass}]^{-2}$. Notice that the Schwinger phase cancels in obtaining the gap equation and therefore we omit it hereafter.
Plugging Eq.~(\ref{eq:prop}) into Eq.~(\ref{eq:condensate}) and taking
the trace, we obtain the chiral condensate for a single quark flavor,
\begin{eqnarray}
-\braket{\bar{\psi}\psi}&=&\nonumber\\
&&\hspace{-15mm}4N_{c}m\int\frac{d^{4}p}{(2\pi)^{4}}\int_{\Lambda}^{\infty}ds\,e^{is\left(p_{\parallel}^{2}-p_{\perp}^{2}\frac{\tan(q_{f}Bs)}{q_{f}Bs}-m^{2}\right)},\label{eq:condensateM}
\end{eqnarray}
where $N_c=3$ is the number of colors.
 Performing the Gaussian integrals over the perpendicular components of momentum and upon Wick rotating to Euclidean space through the replacement $p_{0}\longrightarrow ip_{0}$,
we obtain
\begin{equation}
-\braket{\bar{\psi}\psi}=\frac{N_{c}mq_{f}B}{\pi}\int\frac{d^{2}p_{\parallel}}{(2\pi)^{2}}\int_{\Lambda}^{\infty}ds\,\frac{e^{-is\left(p_{\parallel}^{2}+m^{2}\right)}}{\tan(q_{f}Bs)}.
\end{equation}
Now, in order to take into account thermal effects, we use the Matsubara formalism~\cite{kapusta} which requires the replacements
\begin{equation}
\int_{-\infty}^\infty\frac{dp_{0}}{2\pi} f(p_0)\quad\longrightarrow \quad T\,\sum_{n=-\infty}^{\infty} f(\omega_n),
\end{equation}
where $\omega_{n}=2(n+\nicefrac{1}{2})\pi T$ are the Matsubara frequencies.
On carrying out the remaining Gaussian momentum integral and
performing the change of variable $s\to-is$, we obtain 
\begin{eqnarray}
-\braket{\bar{\psi}\psi} & =&\frac{N_{c}mq_{f}B}{2\pi^{\nicefrac{3}{2}}}\int_\Lambda^\infty\frac{ds}{s^{\nicefrac{1}{2}}}\frac{e^{-sm^{2}}}{\tanh(q_{f}Bs)}
\nonumber \\&& \times 
 T\sum_{n=-\infty}^{\infty}e^{-s\omega_{n}^{2}}.\label{eq:fullcondensate}
\end{eqnarray}
To simplify Eq.~(\ref{eq:fullcondensate}) further, we recall the definition of the third Jacobi Elliptic theta function
\begin{equation}
\Theta_{3}(z,\tau)=1+2\sum_{n=1}^{\infty}q^{n^{2}}\cos(2nz)\;,
\end{equation}
with $q=e^{i\pi\tau}$. In this way,
\begin{eqnarray}   
\sum_{n=-\infty}^{\infty}e^{-s\omega_{n}^{2}}&=&e^{-\pi^2T^2s}\Theta_{3}(i2\pi^2T^2s,i4\pi T^{2}s),\nonumber\\
&=&(4\pi T^{2}s)^{-\frac{1}{2}}\Theta_{3}\left(-\frac{\pi}{2},\frac{i}{4\pi T^{2}s}\right),
\end{eqnarray}
where we have used the inversion formula
\begin{equation}
\Theta_{3}(\tau,z)=(-i\tau)^{-\frac{1}{2}}e^{-\frac{iz^{2}}{\pi\tau}}\Theta_{3}\left(-\frac{z}{\tau},-\frac{1}{\tau}\right).
\end{equation}

Thus, the
chiral condensate simplifies to
\begin{eqnarray}
-\braket{\bar{\psi}\psi} &=&\frac{N_{c}mq_{f}B}{2\pi^{\nicefrac{3}{2}}}\frac{1}{(4\pi)^{\frac{1}{2}}}\int_\Lambda^\infty\frac{ds}{s}\frac{e^{-sm^{2}}}{\tanh(q_{f}Bs)}\nonumber \\
 &&\times\Theta_{3}\left(-\frac{\pi}{2},\frac{i}{4\pi T^{2}s}\right),\label{eq:condensatewtheta}
\end{eqnarray}
and correspondingly, the gap Eq.~(\ref{eq:gapgral}) becomes
\begin{eqnarray}
m & =& m_{q}+GN_{c}\frac{mq_{f}B}{2\pi^{2}}\int_{\Lambda}^{\infty}\frac{ds}{s}\frac{e^{-sm^{2}}}{\tanh(q_{f}Bs)}\nonumber \\
 && \times\Theta_{3}\left(-\frac{\pi}{2},\frac{i}{4\pi T^{2}s}\right).\label{eq:esgapcompleta}
\end{eqnarray}
Next, to explicitly isolate the vacuum from medium contributions, 

we split the first term of the sum 
\begin{equation}
\Theta_{3}\left(-\frac{\pi}{2},\frac{i}{4\pi T^{2}s}\right)=1+2\sum_{n=1}^{\infty}(-1)^{n}e^{-\frac{n^{2}}{4T^{2}s}},
\end{equation}
and hence, adding and subtracting a factor 1 to account for the would-be divergent term to cancel the ultraviolet contribution of the medium~\cite{allofus1,allofus2},  we arrive at

\begin{eqnarray}
m & =&m_{q}+GN_{c}\frac{m}{2\pi^{2}}\Bigg\{ \int_{\Lambda}^{\infty}\frac{ds}{s^{2}}e^{-sm^{2}}\nonumber \\
 &&\hspace{-8mm}+ \int_{0}^{\infty}\frac{ds}{s^{2}}e^{-sm^{2}}\left(\frac{q_{f}Bs}{\tanh(q_{f}Bs)}-1\right)\nonumber \\
 &&\hspace{-8mm}+ 2q_{f}B\int_{0}^{\infty}\frac{ds}{s}\frac{e^{-sm^{2}}}{\tanh(q_{f}Bs)}\sum_{n=1}^{\infty}(-1)^{n}e^{-\frac{n^{2}}{4T^{2}s}}\Bigg\} .
 \label{eq:gapsplit}
\end{eqnarray}
The first integral in Eq.~(\ref{eq:gapsplit}) is the
vacuum term, whereas the second and third terms are the thermomagnetic contribution. The purely thermal contribution is straightforwardly obtained taking the limit $B\to 0$, whereas the purely magnetic contribution comes from the limit $T\to 0$. Notice that the regularization parameter $\Lambda$ remains in the vacuum integral only. On physical grounds, we expect this to be the case as the medium contribution is strongly suppressed in the ultraviolet. Below, we analyze the above equation to explore the parameter space domains where chiral symmetry breaking is possible in the model. 

\section{Critical curves}\label{sec:surf}

\begin{figure}[t!]
\centering{}\includegraphics[width=0.9\columnwidth]{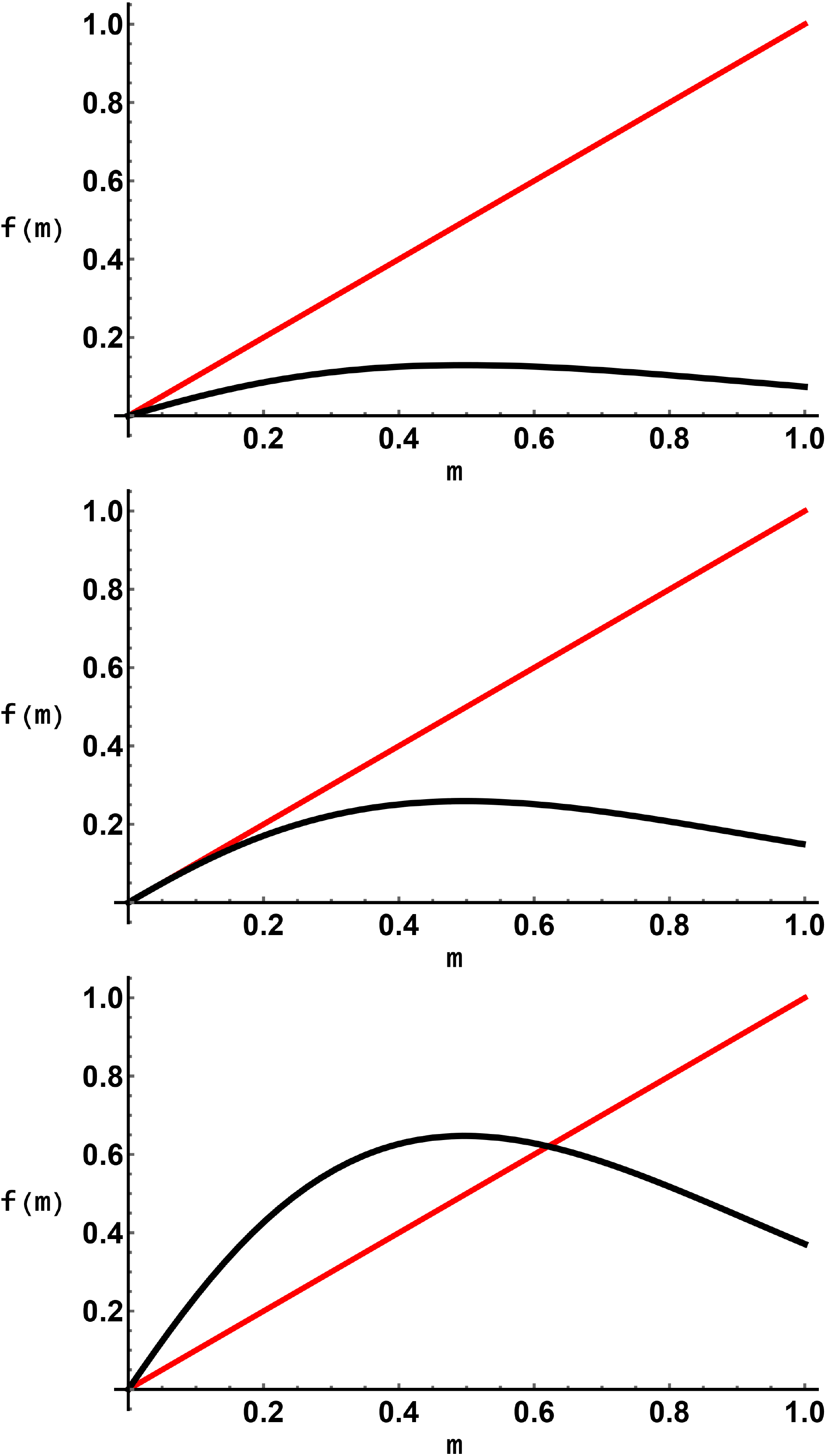}\caption{\label{fig:crit}Function $f(m)$ in eq.~(\ref{rhs}), corresponding to the r.h.s. of the gap equation, as a function of the dynamical mass $m$ for various values of the coupling (in black). Intersections with the (red) line $y=m$ give the solutions. There could be none except for the trivial one $m=0$ (upper panel), one $m\ne 0\ll 1$ (mid panel) or trivial and nontrivial solutions (lower panel) if $G>G_c$, $G=G_c$ or $G>G_c$, respectively.}
\end{figure}

In this section we are interested in deriving the critical curves, in parameter space, that distinguish domains  supporting and/or prohibiting chiral symmetry to be broken.  We start considering the vacuum term alone. From Eq.~(\ref{eq:gapsplit}), setting $T=B=0$ we have
\begin{eqnarray}
m & =&m_{q}+GN_{c}\frac{m}{2\pi^{2}} \int_{\Lambda}^{\infty}\frac{ds}{s^{2}}e^{-sm^{2}}.\label{gap}
\end{eqnarray}
We look for nontrivial solutions $m\ne 0$ to the above expression. In the chiral limit, it is equivalent to find the intersections of the curves
\begin{equation}
    f(m)=GN_{c}\frac{m}{2\pi^{2}} \int_{\Lambda}^{\infty}\frac{ds}{s^{2}}e^{-sm^{2}}\label{rhs}
\end{equation}
as a function of $m$ for all other parameters fixed and the line $y=m$. Depending upon the strength of the coupling constant (see Fig.~\ref{fig:crit}), there could be no intersection whatsoever if the coupling is weak, except for the trivial $m=0$. There exists, however, a critical value $G_c$ where there is one more intersection and correspondingly, a nontrivial solution $m\ne 0$ bifurcates away from the trivial one $m=0$, and for $G>G_c$ at least these two intersections are observed.
Then, to find the critical coupling $G_{c}$ that allows  chiral symmetry breaking, we derive the gap equation~(\ref{gap}) with respect to the mass $m$ and evaluate at $m=0$. This procedure indicates exactly where the trivial and nontrivial solutions  bifurcate from one another  and specifies the value of the coupling $G_c$ required for that purpose. In our case, we have
\begin{eqnarray}
1&=&G_{c}\frac{3}{2\pi^{2}}\int_{\Lambda}^{\infty}\frac{ds}{s^{2}}
\equiv\frac{G_c}{\tilde{\Lambda}},
\end{eqnarray}
where $\tilde{\Lambda}= 2\pi^2\Lambda/3$.
Therefore

$G_{c}=\tilde{\Lambda}$

is the critical value of the coupling above which chiral symmetry is broken in the model. 

\begin{figure}[t!]
\centering{}\includegraphics[width=\linewidth]{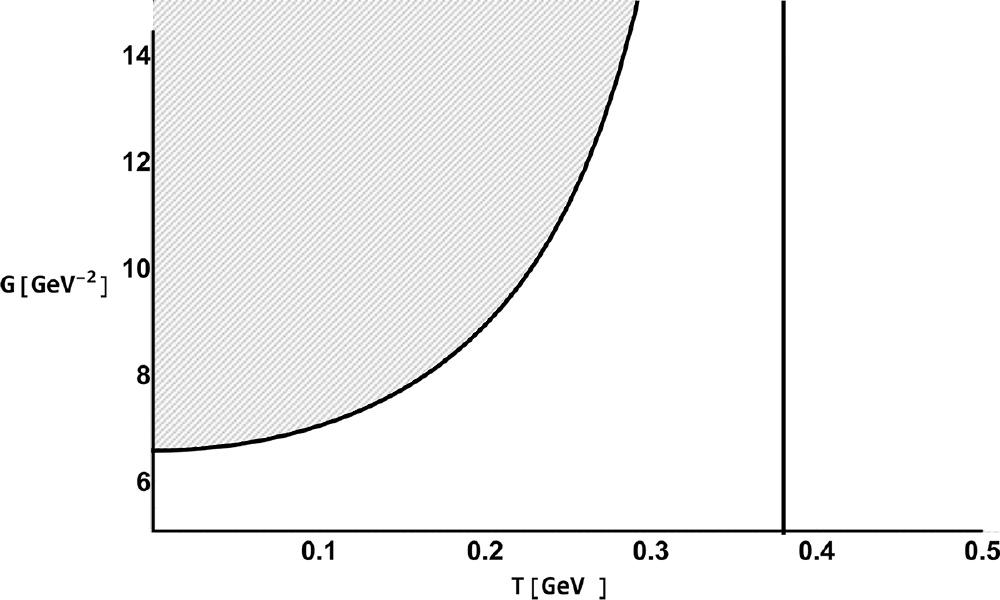}\caption{\label{fig:Critical-coupling-constant T}Coupling constant $G$
 in a thermal bath as a function of $T$. The solid curve corresponds to $G_c^T$ in Eq.~(\ref{eq:G_cT}). 
 Notice that for $T=T_c\equiv 1/\sqrt{G_c}$,
$G_{c}^T$ is divergent. This means that, no matter how strong the coupling constant is, there is no chiral symmetry breaking. Also, for any value of the coupling $G>G_{c}^T$, quark masses are dynamically generated. The scale of the plot is set by $\Lambda=1{\rm GeV}^{-2}$.}
\end{figure}

Next, we consider the effect of a heat bath at temperature $T$. Taking the limit $B\to 0$ in Eq.~(\ref{eq:gapsplit}), on deriving with respect to $m$ and setting $m=0$,  we reach at the critical relation 
\begin{equation}
1= \frac{3}{2\pi^{2}}G_{c}^T\left(\frac{1}{\Lambda}-\frac{2}{3}\pi^{2}T^{2}\right),
\label{eq:0.3}
\end{equation}
where $G_c^T$ stands for the critical coupling required to break chiral symmetry in a heat bath at temperature $T$. Observe that at this point, $G_c^T$  could have a non-trivial dependence of the plasma parameters and thus Eq.~(\ref{eq:0.3}) becomes a self-consistent relation for the critical temperature for chiral symmetry restoration. Assuming that the coupling constant is independent of the temperature, by writing in the form

\begin{equation}
G_{c}^T=\frac{G_c}{1-G_cT^2},
\label{eq:G_cT}
\end{equation}
we notice that at $T=0$ we recover the vacuum limit explicitly. Furthermore, there exist
a critical temperature $T_{c}=1/\sqrt{\tilde{\Lambda}}=1/\sqrt{G_c}$ 
where $G_c^T$ diverges, which means that no matter how strong the coupling is, it is not enough to break chiral symmetry. 
A plot  of $G$  as a function of $T$ for a fixed value of $G_c$ is shown in Fig.~\ref{fig:Critical-coupling-constant T}.
The critical curve $G_{c}^T$ limits the values of $G$ that generate mass from those that do not. Any value of $G$ above the curve of $G_{c}^T$  suffices to generate a dynamical quark mass $m\ne 0$. Thus, the shaded region corresponds to the chirally broken region in parameter space. The vertical line corresponds to the position of $T_c$ and to the right of such a line, chiral symmetry can never be broken. For the general case of a $T$-dependent coupling, of course, Eq.~(\ref{eq:G_cT}) becomes a transcendental, self-consistent equation to find the critical temperatures. The shape of the boundary is expected to be refined accordingly

\begin{figure}
\begin{centering}
\includegraphics[width=\linewidth]{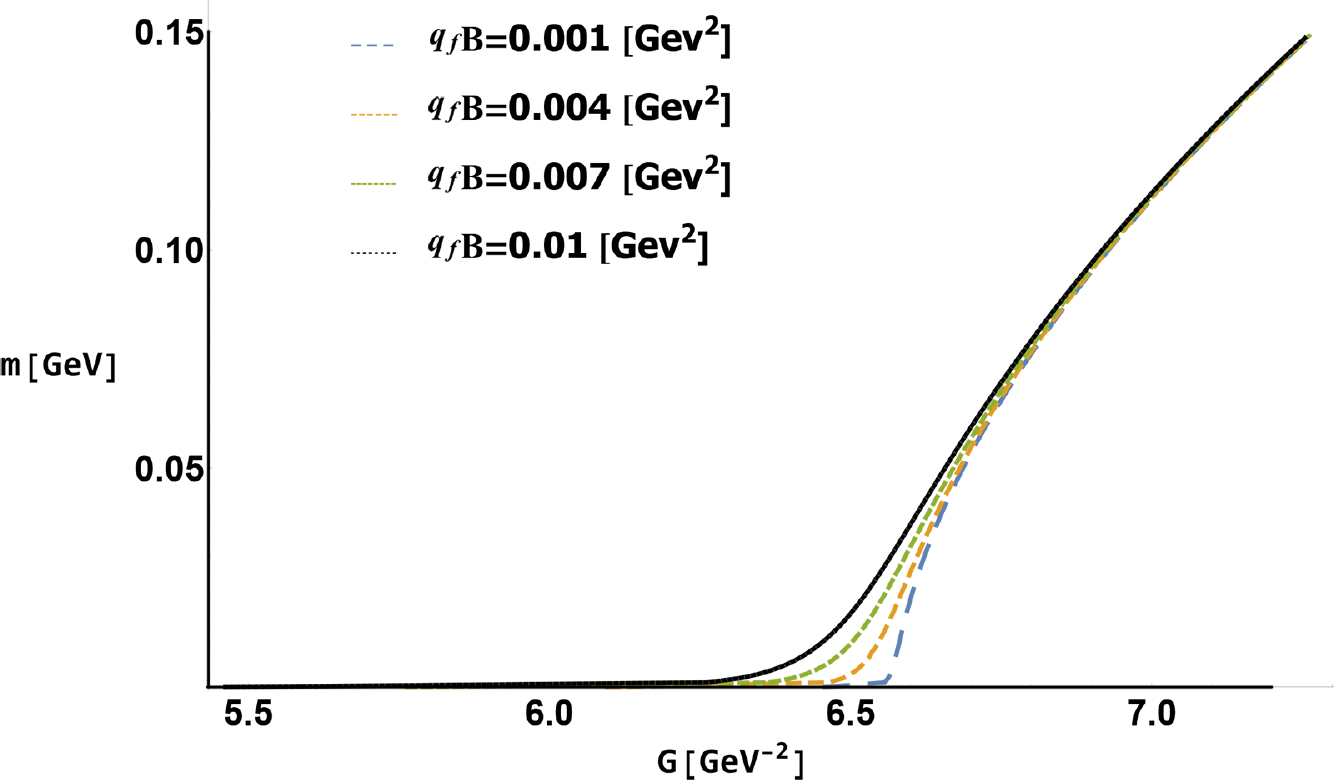}
\par\end{centering}
\caption{\label{fig:m de G magnetico}Dynamically  generated mass $m$ as a function of the coupling $G$ for various values of the magnetic field at fixed $\tilde{\Lambda}=1\,{\rm GeV}^{-2}$.
}
\end{figure}

\begin{figure}
\begin{centering}
\includegraphics[width=\linewidth]{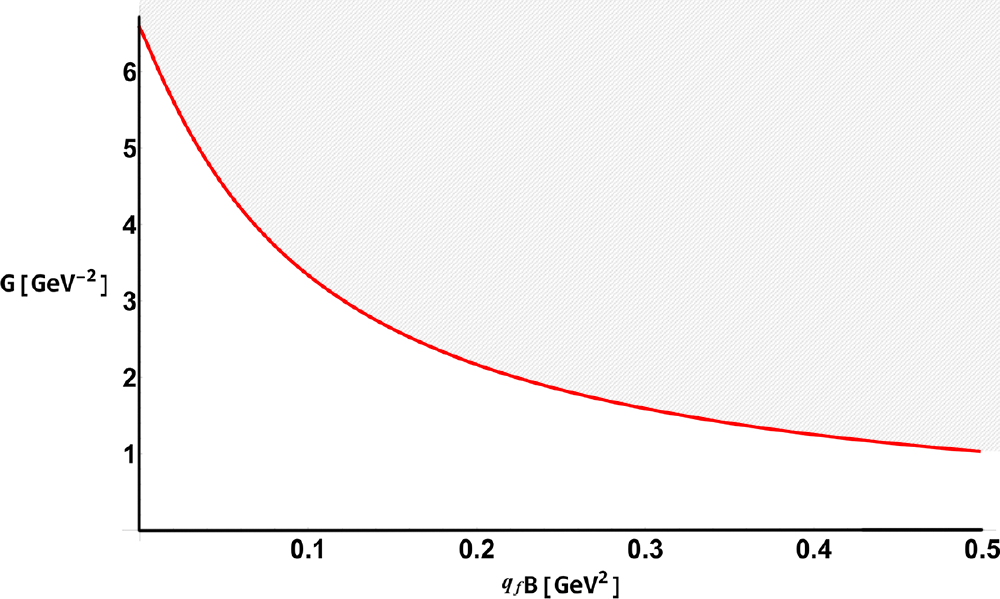}
\par\end{centering}
\caption{\label{fig:puntocriticosuave} Pseudo-critical coupling  as a function of the magnetic field strength. Under the critical curve $\tilde{G}_{c}^M$, shown as a red, solid curve, the generated mass is smaller than $10^{-4}~{\rm GeV}$.
Notice that $\tilde{G}_{c}^M$ gets smaller as the magnetic field strength increases, in accordance with the phenomenon of MC. }

\end{figure}

The influence of solely a magnetic field can be considered from the gap Eq.~(\ref{eq:gapsplit}) in the limit $T\to 0$. Again, differentiating with respect to the mass and setting afterward $m=0$, we obtain
\begin{equation}
1=  \frac{3}{2\pi^{2}}G_{c}^M\left[\frac{1}{\Lambda}+\int_{0}^{\infty}\frac{ds}{s^{2}}\left(\frac{q_fBs}{\tanh(q_fBs)}-1\right)\right],\label{eq:0.4}
\end{equation}
where $G_c^M$ represents the critical coupling needed to generate masses in a magnetized medium for a single quark flavor of electric charge $q_f$. For a magnetic field of arbitrary strength, the integral on the r.h.s. of Eq.~(\ref{eq:0.4}) diverges and thus  $G_{c}^M$  cannot be  defined if we assume the coupling is not dressed by the magnetic field in a nontrivial manner. This circumstance implies that there is generation of masses for any finite value of $G_c^M$ regardless of the magnetic field strength in the mean field limit. Even for weak magnetic fields, a given value of $q_fB$ makes any value of the weak coupling strong enough to form the condensate. This
behavior is reminiscent of the catalytic effect of a magnetic field at zero temperature that promotes the generation of mass through the formation of the chiral condensate. An important observation is that in this regime, the generated mass is extremely small, as shown in  Fig.~\ref{fig:m de G magnetico}.
Nevertheless, by demanding the generated mass to be larger than the current quark masses, namely of ${\cal O}(10^{-4}~{\rm GeV})$, we can define a pseudo-critical coupling $\tilde{G}_{c}^M$ as the coupling needed to generate such a mass. In the weak field regime, we sketch the $G-q_fB$ plane for dynamical generation of quark mass. It is shown as a function $q_fB$ in  Fig.~\ref{fig:puntocriticosuave}. The shaded region correspond to masses larger than the current quark masses and the red (solid) line is the pseudo-critical curve. For a fully dressed coupling, of course, Eq.~(\ref{eq:0.4}) becomes a self consistent relation for the magnetic field needed to break chiral symmetry.

\begin{figure}
\begin{centering}
\includegraphics[width=\linewidth]{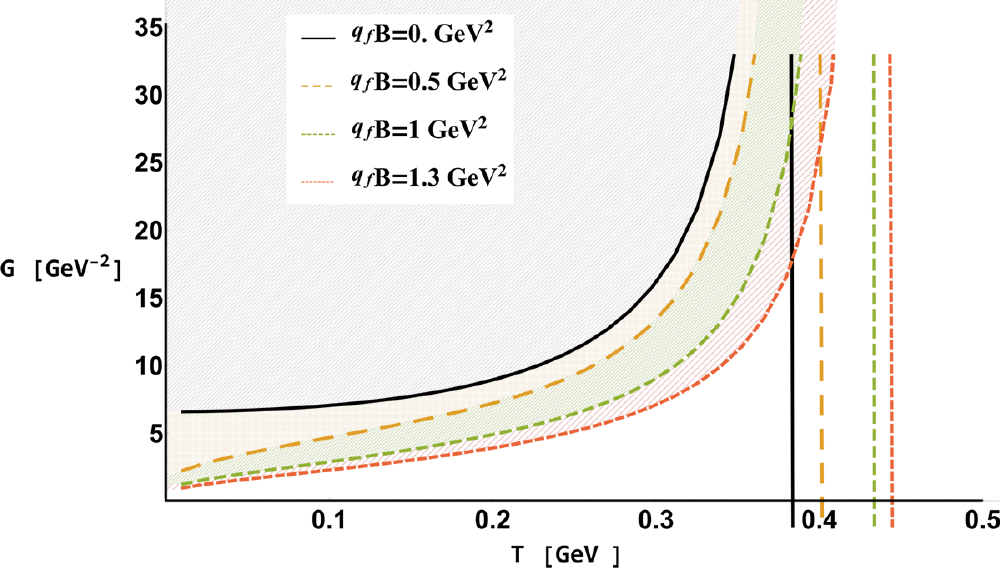}
\par\end{centering}
\caption{\label{fig:Gc critica B y T}Coupling  constant 
as a function of the temperature $T$ for various values of the magnetic
field $q_fB$. The corresponding critical curves $G_{c}^{TM}$ are shown. These separate the chiral assymetric from the symmetric domains. We can see that, even in the presence of the magnetic field, we have a critical temperature $T_{c}$. Also notice that as the magnetic field increases, the critical temperature $T_{c}$ tends to increase. }

\end{figure}

The last case at hand is the full gap equation in a thermomagnetic plasma. In this case, the condition for criticality reads
\begin{eqnarray}
1&= & \frac{3}{2\pi^{2}}G_{c}^{TM}\Bigg[\frac{1}{\Lambda}+\int_{0}^{\infty}\frac{ds}{s^{2}}\left( \frac{q_fBs}{\tanh(q_fBs)}-1\right) \nonumber \\
& & +2q_fB\sum_{n=1}^{\infty}\intop_{0}^{\infty}\frac{ds}{s}\frac{(-1)^{n}\exp(-sn^{2})}{\tanh(\frac{q_fB}{4T^{2}s})}\Bigg],\label{eq:0.5}
\end{eqnarray}
where $G_c^{TM}$ represents the critical coupling in the thermomagnetic medium for a single quark flavor of charge $q_f$ to obtain a mass $m\ne 0$.
The first thing we can readily verify is that the integrals in the r.h.s. of
Eq.~(\ref{eq:0.5}) are convergent. This is so because, unlike the pure magnetic
field case, the leading term of the second integral ($n=1$ in the sum) in the limit $B\to 0$ cancels the divergence of the second term (pure magnetic contribution) in this limit. Physically, we expect such a cancellation because the temperature tries to dissolve the condensate. Thus, there is a
competition between the magnetic field and the temperature to promote and inhibit the generation of a quark mass which leads to
the existence of a critical coupling constant $G_{c}^{TM}$. In Fig.~\ref{fig:Gc critica B y T} we plot the coupling constant  as a function of the
temperature for different values of $q_fB$. Different lines  correspond to the corresponding critical curves $G_{c}^{TM}$. Shaded regions are the chirally asymmetric domains. 
For low temperatures, we observe that the largest value of $G_c^{TM}$ that hits the vertical axis corresponds to the zero magnetic field case, and that such a height diminishes as the strength of the magnetic field increases. This is expected under the view of the MC phenomenon. For larger values of $T$,
vertical lines in the plot correspond to the values of $T$ where each $G_{c}^{TM}$ diverges. In other words, these correspond to the critical temperature above which chiral symmetry can no longer be broken. As the magnetic field increases in strength, the critical temperatures move toward larger values.

\begin{figure}
\begin{centering}
\includegraphics[width=\linewidth]{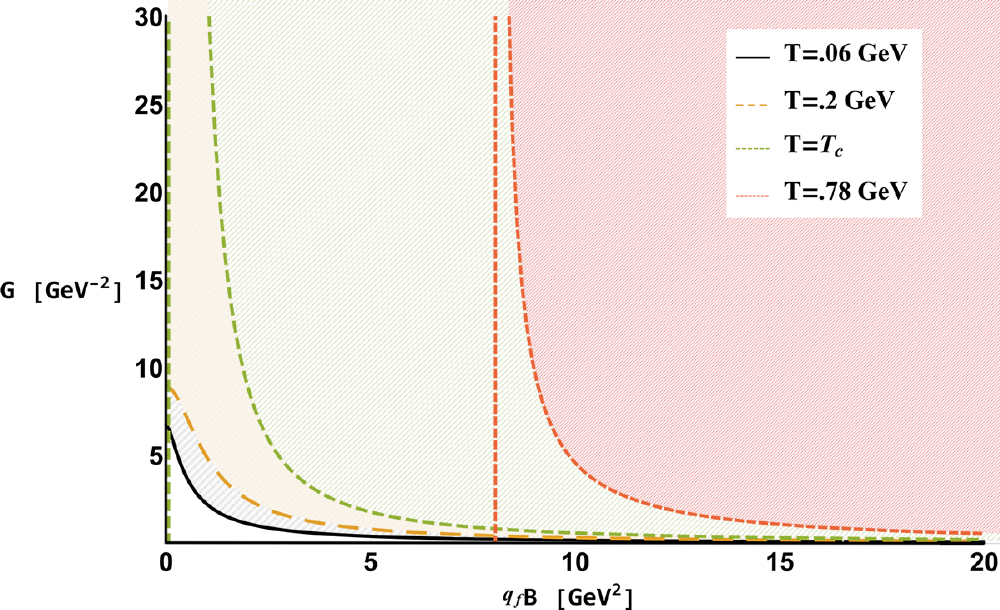}
\par\end{centering}
\caption{\label{fig:critico mag temp} Coupling  constant  as a function of the magnetic field $q_fB$ for various values of the temperature
$T$. Critical curves $G_c^{TM}$ are also shown. For temperatures $T<T_c=1/\sqrt{G_c}$,  the domains of chiral symmetry breaking include any values of $q_fB$. Nevertheless, for $T>Tc$,  there is as critical value of the magnetic
field strength $q_fB_{c}$  such that if the magnetic field $q_fB$ is larger, then chiral symmetry is broken.}
\end{figure}

In Fig.~\ref{fig:critico mag temp} we plot the same
 coupling  as a function of the magnetic field for
various values of the temperature. $G_{c}^{TM}$ curves are also shown and the shaded regions correspond to chirally asymmetric domains. We notice that when temperatures $T<1/\sqrt{G_c}$,  masses are generated for arbitrary values of $q_fB$. Nevertheless, for $T=1/\sqrt{G_c}$ a turnover behavior develops such that for $T>1/\sqrt{G_c}$ a critical strength $q_fB_c$ is required in order to break chiral symmetry. The critical $q_fB_c$ is a vertical line that corresponds also to $G_c^{TM}$. 
Thus, we observe two entirely different behaviors of coupling which separate at $T=T_0$: on the one hand ($T>T_0$), we have always the possibility for chiral symmetry breaking for arbitrary magnetic field strength, and in the other hand ($T>T_0$), a critical magnetic field strength is needed for that purpose. It is therefore naturally expected that for a coupling that is fully dressed by the plasma parameters, this transition between the two regimes is smooth, as observed in the IMC effect. In this direction, below we revise the implications of the criticality condition~(\ref{eq:0.5}) on the magnetized phase diagram.

\section{Inverse magnetic catalysis models of the coupling constant}\label{sec:comp}

In this section we address the issue of the critical hypersurface and its shape according to restrictions arising from IMC. For this purpose, we derive the magnetized phase diagram with several proposals for the coupling from Eq.~(\ref{eq:0.5}). In particular, we test some models which include or not specific dependence of the plasma parameteres in $G$ in accordance with the IMC phenomenon. Explicitly, we consider the following examples
\begin{itemize}
    \item Mean field coupling, $G^{TM}\equiv G^0$, namely, the coupling is independent of the plasma parameters.
    \item Running coupling of QCD in a background magnetic field~\cite{shov}
\begin{equation}
G^{TM}=\frac{G^{0}}{\ln{\bigg{(}e+\frac{|q_fB|}{\Lambda_{QCD}^2}\bigg{)}}},\label{modelolog}
\end{equation}
were $\Lambda_{QCD}^2=300\thinspace {\rm MeV}$ and $G^{0}$ is the value of the coupling constant in vacuum. Here, no explicit dependence to the temperature is considered.
\item A Pad\`e fit~\cite{pade,aftab,costa}
\begin{equation}
G^{TM}(q_fB)=G^0\bigg{(}\frac{1+a\zeta^2+b\zeta^3}{1+c\zeta^2+d\zeta^4}\bigg{)},\label{modeloaftad}
\end{equation}
where $a=0.0108805$, $b=-1.0133\times10^{-4}$, $c=0.02228$, $d=1.84558\times 10^{-4}$,  $\zeta=q_fB/\Lambda_{QCD}^2$ and $\Lambda_{QCD}^2=300\thinspace {\rm MeV}$. This model is temperature independent and is known to reproduce the behavior of the critical temperatures  as a function of $q_fB$ in consistency with lattice results for IMC.

\item A nontrivial fit of the form~\cite{gastao}
\begin{equation}
G^{TM}=c(B)\Bigg[ 1-\frac{1}{1+\exp{[\beta(B)(T_a(B)-T)]}}\Bigg]+s(B)\;,
\label{eq:coupgastao}
\end{equation}
where the parameters $c(B)$, $\beta(B)$, $T_a(B)$ and $s(B)$ are tabulated in Table 1 of Ref.~\cite{gastao}. Assuming the dynamically generated mass to be independent of the plasma parameters, this {\em ansatz} renders the behavior of the critical temperature $T$ vs. $q_fB$ in accordance with lattice simulations for the IMC.

\item A numerical fit suggested in Ref.~\cite{allofus1}, based on reverse engineering of lattice results to derive the behavior of the chiral condensate. The mass function and the coupling in this case have a non-trivial dependence of the plasma parameters.

\end{itemize}

\begin{figure}
\begin{centering}
\includegraphics[width=\linewidth]{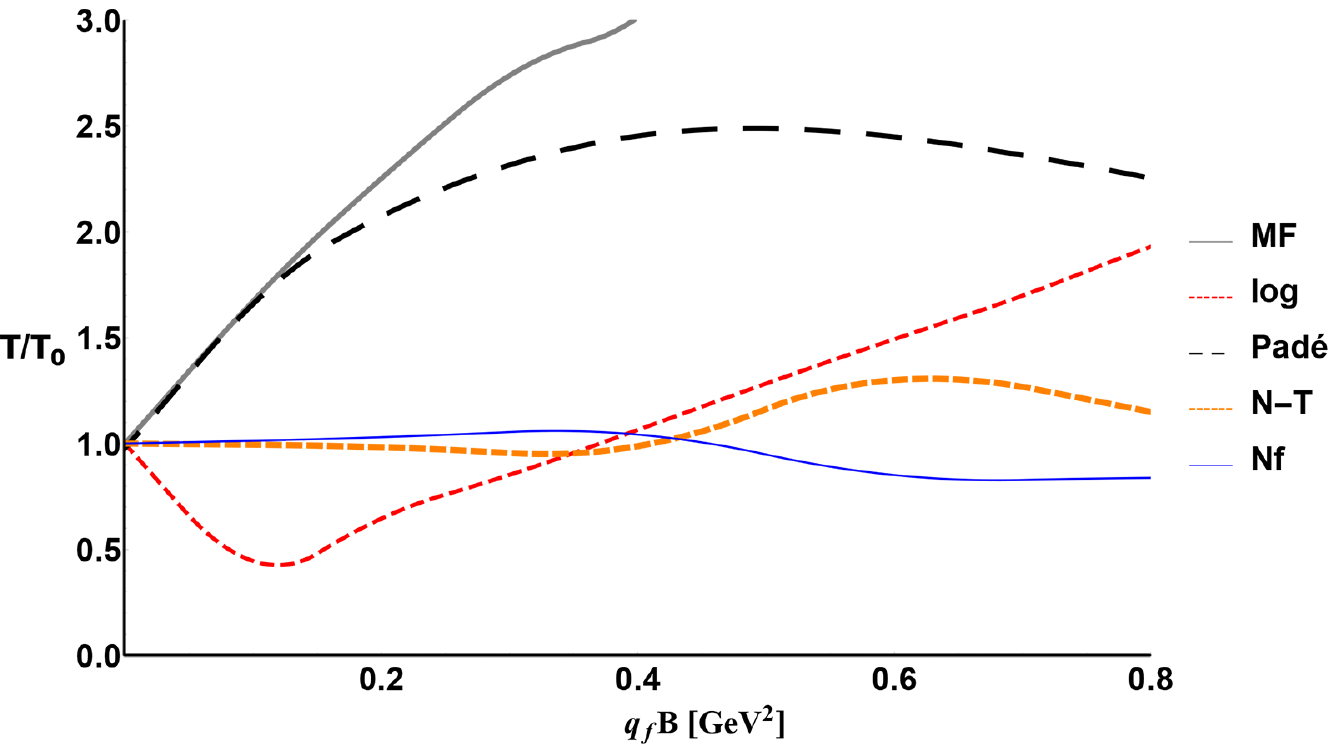}
\par\end{centering}
\caption{\label{fig:coupB} Magnetized phase diagram from the criticality condition~(\ref{eq:0.5}) with different proposals of the coupling constant as explained in the text. The temperature axis is normalized to $T_0$, the critical temperature in absence of the magnetic field.}
\end{figure}

In Fig.~\ref{fig:coupB} we draw the corresponding phase diagrams. The vertical axis is normalized to the critical temperature in absence of the magnetic field $T_0$. We observe that the MF 
coupling is consistent with MC, namely, the critical temperature grows monotonically with the magnetic field strength. For the case of the running coupling, critical temperatures start diminishing and then monotonically increase. Such a behavior comes solely from comparing the magnetic field strength to $\Lambda^2_{QCD}$. The remaining three proposals already include the traits of the IMC phenomenon and show the expected decrease of the critical temperature as the magnetic field increases. The Pad\`e fit shows a very smooth signal of IMC. The apparent decrease of $T$ with increasing $q_fB$ for the nontrivial model~\cite{gastao} is due to numerical accuracy and is not related to the behavior of the logarithmic running coupling. The numerical fit, though not much pronounced, shows an increasing and decreasing behavior of $T$ with increasing the magnetic field. Let it be stressed that in the models of Refs.~\cite{allofus1} and~\cite{gastao}, the coupling was obtained for an average condensates of lattice~\cite{bali:2012a,bali:2012b,bali:2014} and hence we multiply by  2 to put all couplings in the same footing for a single quark species.

\section{Concluding Remarks}\label{sec:concl}

In this article we have studied the critical hypersurface in parameter space that separates the domains where chiral symmetry breaking is possible/forbidden in the magnetized NJL model. In vacuum, it is well known that the coupling must exceed a critical value in order to allow for a non-trivial solution of the gap equation. Adding a heat bath, assuming the coupling to have its mean-field character, there exist a critical value of temperature above which no matter how strong the coupling is, it is not possible to break chiral symmetry. For a pure magnetic background, under the same assumptions, chiral symmetry can be broken for any value of the coupling, in accordance with the universal phenomenon of Magnetic Catalysis. Criticality appears only when we demand the generated mass to be larger than the current quark mass. Finally, in a non-trivial thermomagnetic medium, there exist a competition between the temperature and the magnetic field for masses to be dynamically generated. For temperatures lower that the critical temperature in absence of the magnetic field, it is seen that its strength always promotes the breaking of chiral symmetry. Nevertheless, above this critical temperature, the hypersurface develops hard walls such that a critical magnetic field strength is required to strengthen the coupling and generate masses.  This is might be seen as a seed of the IMC effect.
From the critical relation in Eq.~(\ref{eq:0.5}), we have derived the corresponding phase diagram in the $T-q_fB$ plane assuming the coupling is dressed by the plasma. The mean field and running coupling with the magnetic field are solely compatible with MC. When the coupling is non-trivially dressed with the plasma parameters, the critical temperature is no longer a monotonic increasing function of the magnetic field strength, but exhibits a turn-over effect characteristic of IMC.   Extensions of the present reasoning including a non-trivial dependence of the dynamical mass as in Ref.~\cite{allofus1} as well as for non local models are under study. Findings will be reported elsewhere.

\begin{acknowledgements}
We acknowledge valuable discussions with A. Ayala and A. Bashir, M. Loewe, A.J. Mizher and C. Villavicencio. AR acknowledges support from Consejo Nacional de Ciencia y Tecnolog\'ia under grant  256494.
\end{acknowledgements}


\begin{thebibliography}{99}
\bibitem{review} J.~O. Andersen, W.~R. Naylor and A. Tranberg, {\em Rev. Mod. Phys.} {\bf 88}, 025001 (2016).

\bibitem{skokov09} V. Skokov, A.Y. Illarionov, V. Toneev, {\em Int. J. Mod. Phys. A} {\bf 24}, 5925 (2009).

\bibitem{skokov2} L. McLerran and V. Skokov, {\em Nucl. Phys. A} {\bf 929}, 184 (2014).

\bibitem{bali:2012a} G. Bali, F. Bruckmann, G. Endrodi, Z. Fodor, S. Katz, {\em et
al.}, {\em JHEP} {\bf 1202}, 044 (2012).

\bibitem{bali:2012b}  G. Bali, F. Bruckmann, G. Endrodi, Z. Fodor, S. Katz, {\em et al., Phys. Rev. D} {\bf 86}, 071502 (2012).

\bibitem{bali:2014} G. Bali, F. Bruckmann, G. Endrodi, S. Katz, and A. Shafer, JHEP. 1408, 177 (2014).



\bibitem{shovkovy} I.~A. Shovkovy, {\em Lect. Notes Phys.} {\bf 871}  13-49, (2013).


\bibitem{miranskyreview} 
 V. A. Miransky, I. A. Shovkovy, {\em Phys. Rept.} {\bf 576}, 1 (2015).

\bibitem{shov} V. A. Miransky and I. A. Shovkovy, {\em Phys. Rev. D}{\bf 66},
045006 (2002).


\bibitem{ayala-vertex} 	A. Ayala, C.~A. Dominguez, L.~A. Hernandez, M. Loewe and  R. Zamora,  {\em Phys. Lett. B} {\bf 759}, 99 (2016).

\bibitem{ayalaetal:LSM1}  A. Ayala, M. Loewe, Ana Julia Mizher, R. Zamora, {\em Phys. Rev. D} {\bf 90} (2014) no.3, 036001;

\bibitem{ayalaetal:LSM2}
A. Ayala, M. Loewe, R. Zamora, {\em Phys.Rev. D} {\bf 91}  no.1, 016002 (2015);





\bibitem{NJL} Y. Nambu and G. Jona-Lasinio, {\em Phys. Rev.} {\bf 122}, 345,  (1961);\\
Y. Nambu and G. Jona-Lasinio, {\em Phys. Rev.} {\bf 124}, 246, (1961).


\bibitem{klevansky} S. Klevansky, {\em Rev. Mod. Phys.} {\bf 64}, 649 (1992).

\bibitem{buballa} M. Buballa, {\em Phys. Rept.} {\bf 407}, 2015 (2005)

\bibitem{PNJL} K. Fukushima, {\em Phys. Lett. B}{\bf 591}, 277 (2004);  \\
E. Megias, E. Ruiz Arriola, and L.L. Salcedo, {\em Phys. Rev. D} {\bf 74}, 065005 (2006); \\
E. Megias, E. Ruiz Arriola, and L.L. Salcedo, {\em Phys. Rev. D} {\bf 74}, 114014 (2006); \\
M. Ciminale, R. Gatto, N. D. Ippolito, G. Nardulli, and M. Ruggieri,
{\em Phys. Rev. D}{\bf 77}, 054023 (2008);\\
C. Ratti, M. A. Thaler, W. Weise, {\em Phys. Rev. D} {\bf 73}, 014019 (2006);\\
H.-M. Tsai and B. M\"uller, {\em J. Phys. G: Nucl. Part. Phys.} {\bf 36}, 075101 (2009)  .

\bibitem{nNJL} 
M. Buballa, S. Krewald, {\em Phys. Lett. B} {\bf 294}, 19 (1992);\\
R. D. Bowler and M. C. Birse, {\em Nucl. Phys. A}{\bf 582}, 655 (1995);\\
R.~S. Plant, M.~C. Birse, {\em Nucl. Phys. A}{\bf 628}, 607 (1998); \\
I. General, D. Gomez Dumm, N. N. Scoccola, 	{\em Phys. Lett. B} {\bf 506}, 267 (2001).


\bibitem{pade} M. Ferreira, P. Costa, O. Louren\c co, T Frederico and C. Provid\^encia, {\em Phys. Rev. D} {\bf 89}, 116011 (2014).
  
\bibitem{aftab} A Ahmad and  A Raya  {\em J. Phys. G} {\bf 43} (2016) no.6, 065002.
  
  
\bibitem{allofus1} A. Ayala , C. A. Dominguez, L.~A. Hernandez, M. Loewe, A. Raya, J.~C. Rojas, C. Villavicencio, {\em Phys. Rev. D} {\bf 94}, 054019 (2016).

\bibitem{allofus2} 
A. Ayala, L.~A. Hernandez, M. Loewe, A. Raya, J.~C. Rojas, R. Zamora,  {\em Phys. Rev. D} {\bf 96} (2017) no.3, 034007. 


\bibitem{gastao} R.~L.~S. Farias, V.~S. Timoteo, S.~S. Avancini, M.~B. Pinto, and Gastao Krein, {\em  Euro. Phys. Jour. A} {\bf 53}(5):101, (2017).

\bibitem{costa} M. Ferreira, P. Costa and C. Provid\^encia, {\em Phys.Rev. D} {\bf 97} (2018) no.1, 014014 

\bibitem{marcelonnjl}  	M. Loewe, F. Marquez, C. Villavicencio  and  R. Zamora,
{\em Int. J. Mod. Phys. A} {\bf 30},  no.21, 1550123 (2015).


\bibitem{norberto} 	D. G\'omez Dumm, M.~F. Izzo Villafa\~ne, S. Noguera, V.~P. Pagura and N.~N. Scoccola, {\em Phys. Rev. D} {\bf 96}  no.11, 114012 (2017).


\bibitem{schwinger} J. Schwinger, Phys. Rev. 82, 664 (1951).

\bibitem{kapusta} J.~I. Kapusta and C. Gale, {\em Finite-temperature field theory. Principles and applications,} 2nd edition (2011).  Cambridge Monographs on Mathematical Physics. ISBN: 9780521173223.

\end{thebibliography}
\end{document}